\documentstyle[twoside,fleqn,espcrc2,epsf]{article}

\def\lsim{\raise0.3ex\hbox{$<$\kern-0.75em\raise-1.1ex\hbox{$\sim$}}}
\def\gsim{\raise0.3ex\hbox{$>$\kern-0.75em\raise-1.1ex\hbox{$\sim$}}}

\newcommand{\pslash}{p\kern-1ex /}
\newcommand{\Dslash}{{\cal D}\kern-1.5ex /}

\newcommand{\J}[4]{{#1} {\bf #2} (#3) #4}

\newcommand{\NP}{Nucl.~Phys.}

\newcommand{\PRL}{Phys.~Rev.~Lett.}

\title{
\vspace*{-2.cm}
\begin{flushright}
{\normalsize UTHEP-452}\\
{\normalsize UTCCP-P-117}\\
\end{flushright}
Non-perturbative renormalization for a renormalization
group improved gauge action
\thanks{Talk presented by K.~Ide.}}

\author{
CP-PACS Collaboration: \\
S.~Aoki\rlap,%
\address{Institute of Physics, University of Tsukuba,
	Tsukuba, Ibaraki 305-8571, Japan}
R.~Burkhalter\rlap,$^{\rm a,}$%
\address{Center for Computational Physics, University of Tsukuba,
	Tsukuba, Ibaraki 305-8577, Japan}
M.~Fukugita\rlap,%
\address{Institute for Cosmic Ray Research, University of Tokyo,
	Kashiwa, Chiba 277-8582, Japan}
S.~Hashimoto\rlap,%
\address{High Energy Accelerator Research Organization (KEK),
	Tsukuba, Ibaraki 305-0801, Japan}
K.~Ide\rlap,$^{\rm a}$
N.~Ishizuka\rlap,$^{\rm a,b}$
Y.~Iwasaki\rlap,$^{\rm a,b}$
K.~Kanaya\rlap,$^{\rm a}$
T.~Kaneko\rlap,$^{\rm d}$
Y.~Kuramashi\rlap,$^{\rm d}$
V.~Lesk\rlap,$^{\rm b}$
M.~Okawa\rlap,$^{\rm d}$
Y.~Taniguchi\rlap,$^{\rm a}$
A.~Ukawa$^{\rm a,b}$ and
T.~Yoshi\'e$^{\rm a,b}$
}

\begin{document}

\pagestyle{empty}


\begin{abstract}

Renormalization constants of vector ($Z_V$) and axial-vector ($Z_A$)
currents are determined non-perturbatively in quenched QCD
for a renormalization group improved gauge action and a tadpole improved clover
quark action using the Schr\"odinger functional method.
Non-perturbative values of $Z_V$ and $Z_A$ turn out to be smaller than 
the one-loop perturbative values by $O(10\%)$ at $a^{-1}\approx 1$~GeV. 
A sizable scaling violation of meson decay constants $f_\pi$ and $f_\rho$ 
observed with the one-loop renormalization factors remains 
even with non-perturbative renormalization.

\end{abstract}

\maketitle

\section{Introduction}

Reliable lattice calculations of hadronic matrix elements and
quark masses require both high precision numerical simulations
and non-perturbative determinations of renormalization constants
($Z$-factors).
The CP-PACS collaboration recently carried out 
a sophisticated spectrum calculation in $N_f=2$ full 
QCD~\cite{ref:CPPACS-NF2} using a renormalization group (RG)
improved gauge action and a tadpole improved clover quark action. 
However, non-perturbative $Z$-factors were not available for 
this action combination. Hence analyses had to rely on one-loop 
perturbative values. 

As a first step toward a systematic study of non-perturbative
renormalization for this action, we apply the Schr\"odinger 
functional method~\cite{ref:SFMethod} to calculations of $Z$ factors 
for vector ($Z_V$) and axial-vector ($Z_A$) currents 
in quenched QCD with the same improved action. 
We examine in particular whether a large scaling violation 
of meson decay constants observed for this action~\cite{ref:CPPACS-NF2}
is improved with non-perturbative $Z$-factors.
We report preliminary results in these proceedings.



\section{Calculational Method}\label{sec:calculation}

We follow the method developed by the ALPHA collaboration~\cite{ref:ALPHA-Z}. 
Namely, we use a lattice geometry of $L^3\cdot T$ with 
$T=2L$ for $Z_V$ with a vector operator at $t=L$, and
$T=3L$ for $Z_A$ with two axial vector operators at
$t=L$ and $t=2L$, except at $\beta=2.2$ and 2.4 for 
$Z_A$ (see sec.~\ref{sec:exceptional} for details of this exception). 
Tree-level values are used for coefficients of boundary counter
terms of the action.
For improving the axial current, we use the one-loop perturbative value
for the coefficient $c_A$.

Values of $Z_V$ and $Z_A$ are determined for $\beta=2.2$ -- 8.0
which almost covers the range of the CP-PACS quenched 
spectrum calculation~\cite{ref:CPPACS-NF2}, $\beta =2.187$ -- 2.575. 
Physical size is normalized at $\beta=$ 2.6 on an $8^3$ lattice.
For other $\beta$ values, two lattice sizes are analyzed
to match the physical size using the string tension.
Our action has $O(a)$ errors since we employ a tadpole improved value 
of $c_{\rm sw}=(1-0.8412/\beta)^{-3/4}$.
Therefore we extrapolate/interpolate results linearly in $1/L$.
We have analyzed 300--4000 configurations depending on $\beta$
value and lattice size. 

\section{Exceptional Configurations}\label{sec:exceptional}

It is straight-forward to calculate $Z$-factors for
$\beta > 2.4$ for $Z_V$ and $\beta > 2.8$ for $Z_A$.
We find reasonable plateaux in the ratio of Green functions 
for the $Z$-factors in spite of the
$O(a)$ error of the action, which implies viability of 
the Schr\"odinger functional method for our action.

\begin{figure}[t]
\centerline{\epsfysize=6.0cm \epsfbox{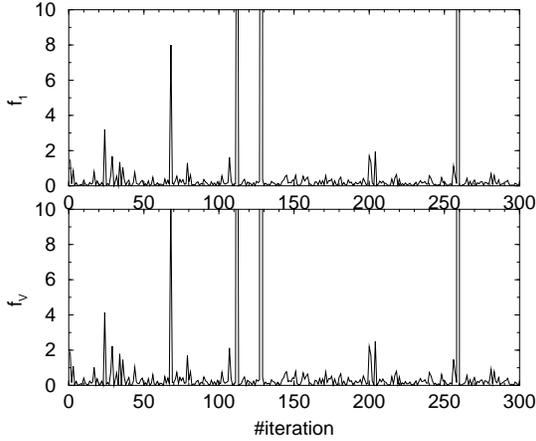}}
\vspace{-1.1cm}
\caption{Time history of $f_1$ and $f_V$ at $\beta=2.4$ on 
an $8^3\times 16$ lattice.}
\label{fig:f1fv}
\vspace{-0.7cm}
\end{figure}

However, for lower $\beta$ values on a large
lattice, anomalously large values appear in the ensemble of 
$f_1$, $f_V$ and $f_{AA}$ where $Z_V={f_1/f_V}$ and $Z_A=\sqrt{f_1/f_{AA}}$.
This is illustrated with a time history of $f_1$ and $f_V$ at $\beta=2.4$ 
in Fig.~\ref{fig:f1fv}.

\begin{figure}[t]
\centerline{\epsfxsize=6.4cm \epsfbox{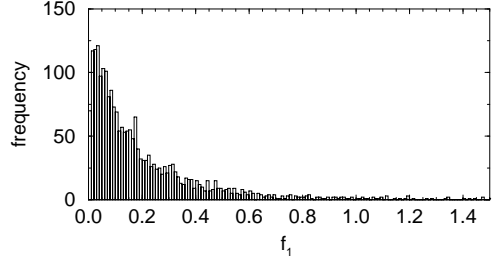}}
\vspace{-1.1cm}
\caption{Histogram of $f_1$ at $\beta=2.4$ on an $8^3\times 16$ lattice.}
\label{fig:hist}
\vspace{-0.8cm}
\end{figure}

\begin{figure}[t]
\centerline{\epsfxsize=6.4cm \epsfbox{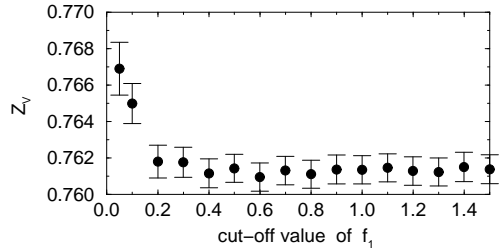}}
\vspace{-1.1cm}
\caption{
Values of $Z_V$ estimated by discarding configurations with
$f_1$ larger than the cutoff value.
Data are for an $8^3\times 16$ lattice at $\beta=2.4$. 
}
\label{fig:cutoff}
\vspace{-0.7cm}
\end{figure}

In order to estimate $Z$-factors at low $\beta$ values,
we have investigated the properties of these ``exceptional configurations''.  
We find :
i) Large values of $f_1$ and $f_V$ for $Z_V$ and $f_1$ and $f_{AA}$ for 
$Z_A$ are strongly correlated (see Fig.~\ref{fig:f1fv}).
ii) Histograms of $f$'s have a long tail toward  very large values 
as shown in Fig.~\ref{fig:hist}.
We then impose a cutoff in taking the average of the $f$'s, and 
find that $Z$-factors are stable against change of the cutoff
as long as anomalously large values are discarded, as 
the numerator and denominator for $Z$-factors are correlated and 
effects mostly cancel out. See Fig.~\ref{fig:cutoff}. 

We then estimate $Z$-factors for low 
$\beta$ values taking a certain value of the cutoff.
For $Z_A$ at $\beta=2.2$ and 2.4, the lattice geometry is also changed from 
$T=3L$ to $T=2L$ because ``exceptional configurations'' appear
very frequently  for the original geometry to the extent that the cutoff 
analysis above does not work. 
We have checked at $\beta=2.6$ that the change of 
geometry does not lead to any significant difference in $Z_A$. 

\section{Results for $Z$-factors}\label{sec:results}

\begin{figure}[t]
\centerline{\epsfxsize=6.4cm \epsfbox{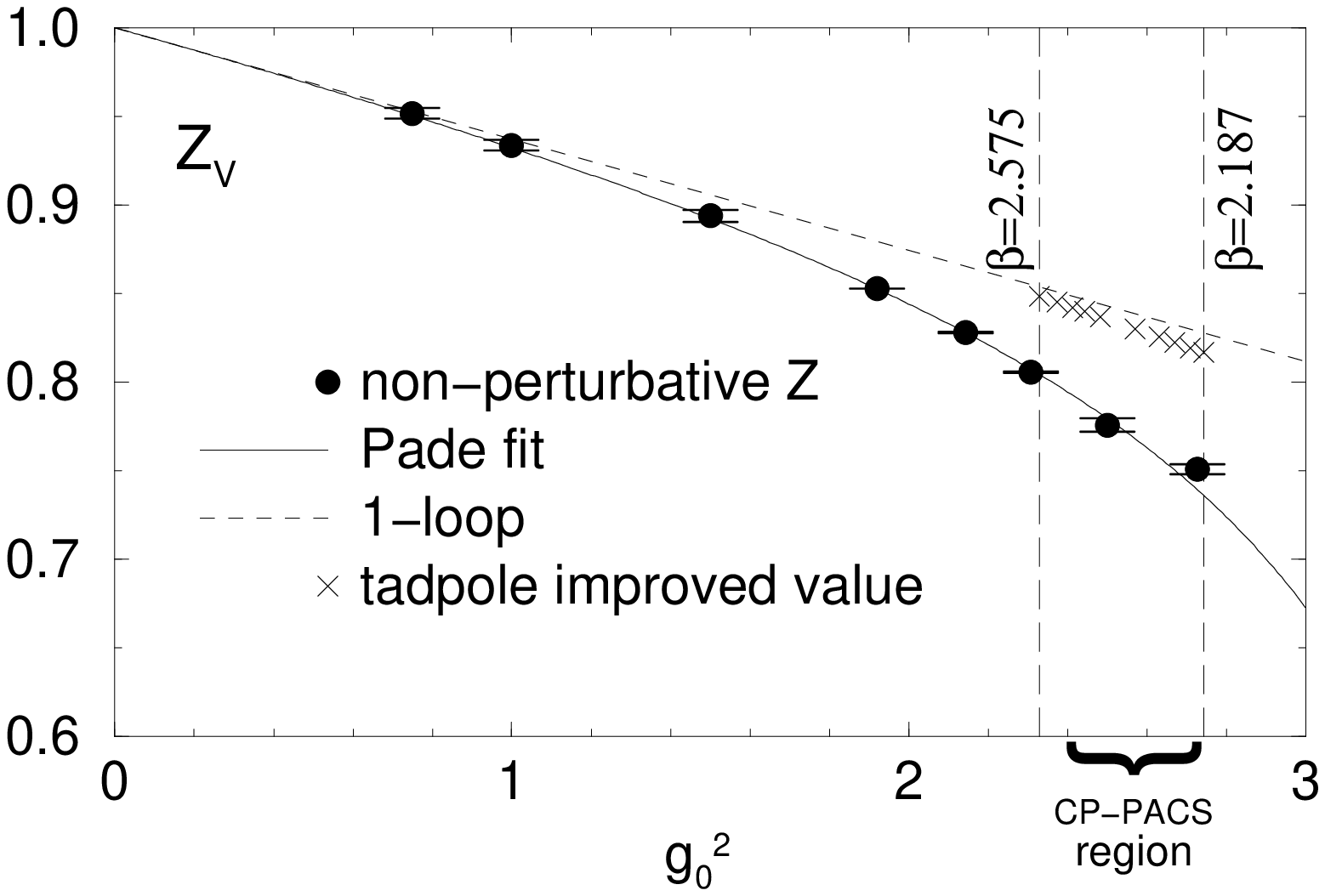}}
\centerline{\epsfxsize=6.4cm \epsfbox{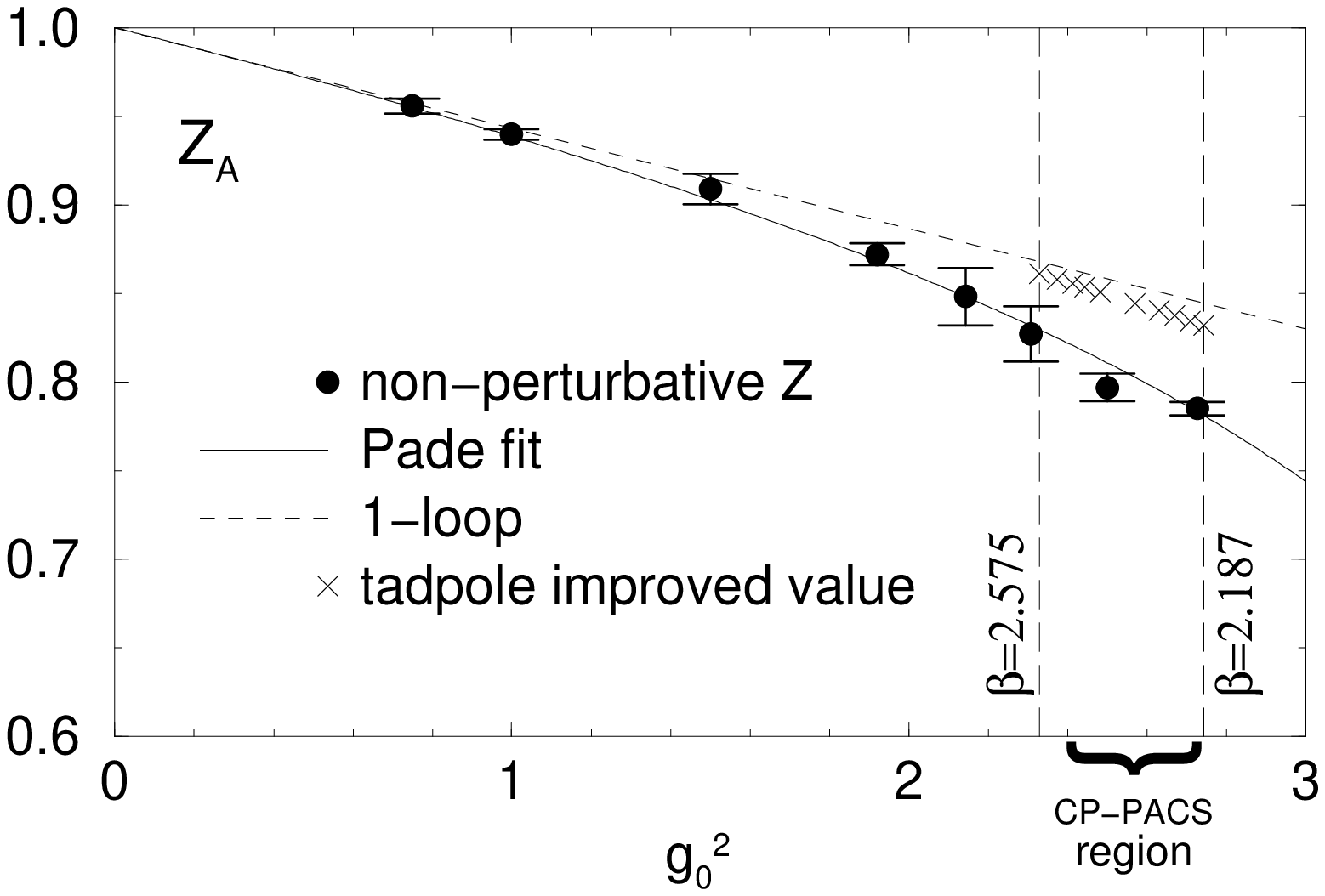}}
\vspace{-1cm}
\caption{Results for $Z_V$ (top) and $Z_A$ (bottom).}
\label{fig:result}
\vspace{-0.5cm}
\end{figure}

In Fig.~\ref{fig:result} we show results of $Z$-factors 
together with Pad\'e fits (solid curves in the figure) to them. 
Non-perturbative estimates give values smaller than the one-loop 
perturbative ones (dashed lines) by about 10 \% (6\%) for $Z_V$ ($Z_A$) 
at the largest coupling of the CP-PACS simulation, $\beta=2.187$.

\section{Scaling Property of Decay Constants}\label{sec:scaling}

\begin{figure}[t]
\centerline{\epsfxsize=6.4cm \epsfbox{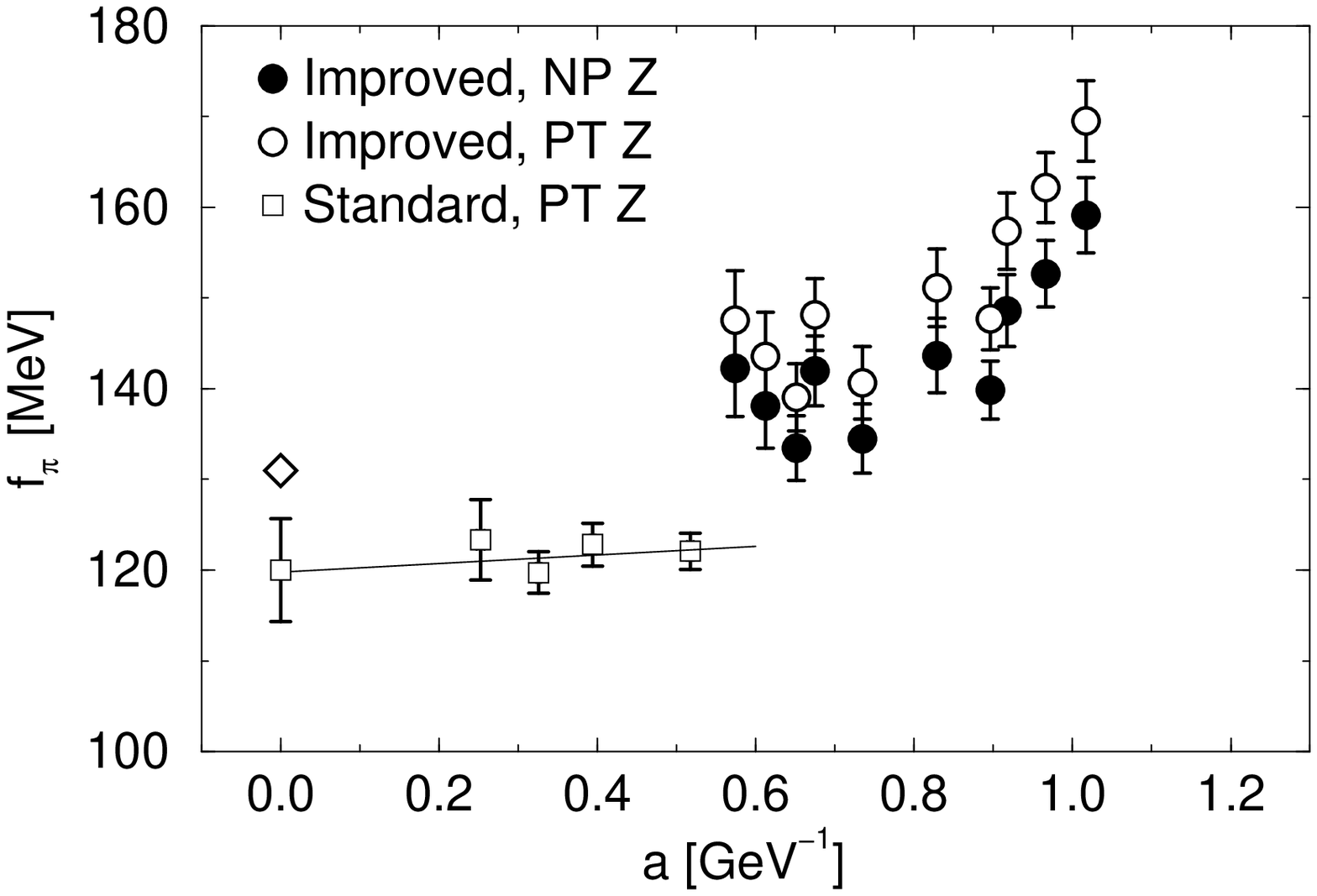}}
\centerline{\epsfxsize=6.4cm \epsfbox{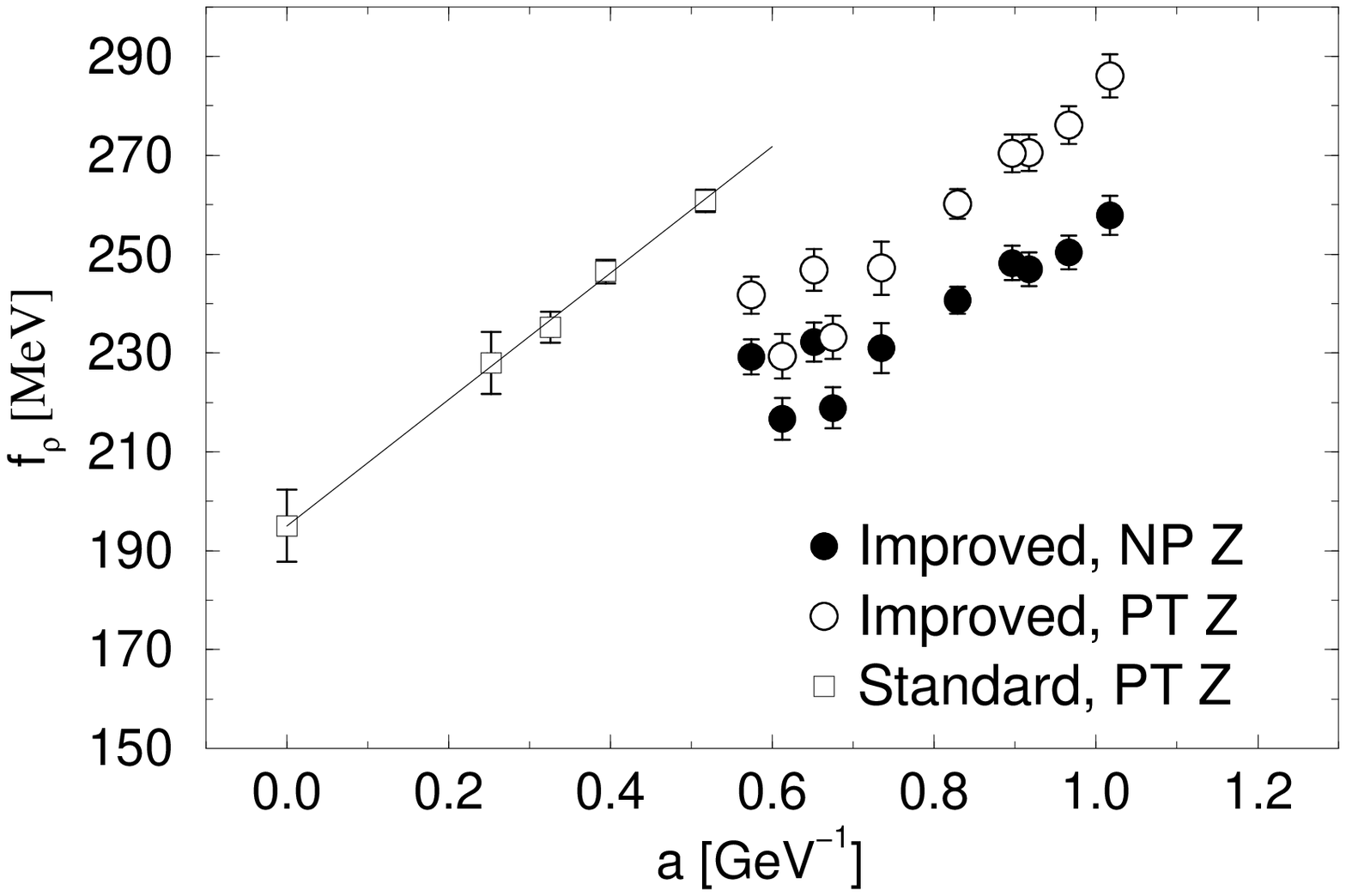}}
\vspace{-1cm}
\caption{$f_\pi$ (top) and $f_\rho$ (bottom) vs. $a$ for
our improved action with non-perturbative (NP) and perturbative
(PT) $Z$-factors together with results for the standard 
action~\protect\cite{ref:CPPACS-quench}.}  
\label{fig:decay}
\vspace{-0.5cm}
\end{figure}

We compare in Fig.~\ref{fig:decay} $f_\pi$ and $f_\rho$
determined with non-perturbative (filled circles) and 
perturbative (open circles) $Z$-factors.  Also shown are 
the results from the standard plaquette and Wilson action
(squares)\cite{ref:CPPACS-quench} using the perturbative 
$Z$-factors.

We observe that, even with the non-per\-tur\-bative $Z$-factors, 
large scaling violation
of meson decay constants remains for the range we have investigated. 
A possible reason is the necessity of 
non-perturbatively fixing the $O(a)$ and perhaps higher terms in the 
currents themselves.  For the axial vector current, it will be 
worth investigating if non-perturbative estimates of the $O(a)$ coefficient 
$c_A$ yield a large value. 

\section{Conclusions}

We have successfully applied the Schr\"odinger functional method
to calculations of $Z_V$ and $Z_A$ for the combination of a RG-improved 
gauge action and a tadpole improved clover quark action down to the lattice 
spacings $a^{-1}=1-2$~GeV where the quenched CP-PACS data for decay constants 
were taken.

While $Z$-factors estimated non-perturbatively are 
smaller by $O(10\%)$ than perturbative ones for this range,  
there still remain large scaling violations of $O(a)$ and higher
in meson decay constants with non-perturbative $Z$-factors.
Further work is needed to examine if hadronic matrix elements could be 
reliably extracted at lattice spacings much coarser than 
$a^{-1}\approx 2$~GeV with operators improved non-perturbatively at $O(a)$ and beyond.

This work is supported in part by Grants-in-Aid of the Ministry of Education 
(Nos.~10640246, 
10640248, 
11640250, 
11640294, 
12014202, 
12304011, 
12640253, 
12740133, 
13640260  
). 
VL is supported by the Research for Future Program of JSPS
(No. JSPS-RFTF 97P01102).

\end{document}